\documentclass[fleqn,usenatbib]{mnras}

\usepackage{newtxtext,newtxmath}
\usepackage{xspace,amsmath}
\usepackage{array,colortbl}

\usepackage[T1]{fontenc}
\usepackage{ae,aecompl}

\newcommand{\hii}{\relax \ifmmode {\mbox H\,{\scshape ii}}\else H\,{\scshape ii}\fi}
\newcommand{\mi}{\relax \ifmmode {\mu{\mbox m}}\else $\mu$m\fi}
\newcommand{\ha}{\relax \ifmmode {\mbox H}\alpha\else H$\alpha$\fi}
\newcommand{\hb}{\relax \ifmmode {\mbox H}\beta\else H$\beta$\fi}

\newcommand{\sii}{\relax \ifmmode {\mbox S\,{\scshape ii}}\else S\,{\scshape ii}\fi}
\newcommand{\siii}{\relax \ifmmode {\mbox S\,{\scshape iii}}\else S\,{\scshape iii}\fi}
\newcommand{\nii}{\relax \ifmmode {\mbox N\,{\scshape ii}}\else N\,{\scshape ii}\fi}
\newcommand{\oi}{\relax \ifmmode {\mbox O\,{\scshape i}}\else O\,{\scshape i}\fi}
\newcommand{\oii}{\relax \ifmmode {\mbox O\,{\scshape ii}}\else O\,{\scshape ii}\fi}
\newcommand{\oiii}{\relax \ifmmode {\mbox O\,{\scshape iii}}\else O\,{\scshape iii}\fi}
\newcommand{\neiii}{\relax \ifmmode {\mbox Ne\,{\scshape iii}}\else Ne\,{\scshape iii}\fi}

\newcommand{\rdostres}{\relax \ifmmode {\,\mbox{R}}_{\rm 23}\else \,\mbox{R}$_{\rm 23}$\fi}

\newcommand{\ciii}{\relax \ifmmode {\mbox O\,{\scshape iii}}\else C\,{\scshape iii}\fi}
\newcommand{\civ}{\relax \ifmmode {\mbox O\,{\scshape iii}}\else C\,{\scshape iv}\fi}
\newcommand{\heii}{\relax \ifmmode {\mbox He\,{\scshape ii}}\else He\,{\scshape ii}\fi}

\usepackage{graphicx}	
\usepackage{amsmath}	

\title[Abundance determination in EELGs]{Extreme Emission-Line Galaxies in SDSS.
I. Empirical and model-based calibrations of chemical abundances}

\author[E. P{\'e}rez-Montero et al.]{
E. P{\'e}rez-Montero$^{1}$\thanks{E-mail: epm@iaa.es (EPM)},
R. Amor\'\i n$^{2,3}$, J. S\'anchez Almeida$^{4,5}$,
J.M. V\'\i lchez$^1$, \and
R. Garc\'\i a-Benito$^1$, C. Kehrig$^1$ \\
\\
$^{1}$Instituto de Astrof\'\i sica de Andaluc\'\i a. CSIC. Apartado de correos 3004. 18080, Granada, Spain.\\
$^{2}$Instituto de Investigaci\'on Multidisciplinar en Ciencia y Tecnolog\'ia, Universidad de La Serena, Raul Bitr\'an 1305, La Serena, Chile. \\
$^{3}$Departamento de Astronom\'ia, Universidad de La Serena, Av. Juan Cisternas 1200 Norte, La Serena, Chile. \\
$^{4}$Instituto de Astrof\'\i sica de Canarias. C/ V\'\i a Lactea s/n. La Laguna, Tenerife, Spain.\\
$^{5}$ Departamento de Astrof\'\i sica, Universidad de La Laguna, Tenerife, Spain. \\
}

\date{Accepted XXX. Received YYY; in original form ZZZ}

\pubyear{2019}

\begin{document}
\label{firstpage}
\pagerange{\pageref{firstpage}--\pageref{lastpage}}
\maketitle

\begin{abstract}
Local star-forming galaxies show properties that are thought to differ from galaxies in
the early Universe. Among them, the ionizing stellar populations and the gas geometry
make the recipes designed to derive chemical abundances from nebular emission lines
to differ from those calibrated in the Local Universe.
A sample of 1969 Extreme Emission Line Galaxies (EELGs) at a redshift 0 $\lesssim z \lesssim$ 0.49, selected from the
{\it Sloan Digital Sky Survey} (SDSS) to be local analogues of high-redshift galaxies,
was used to analyze their most prominent emission lines and to derive
total oxygen abundances and nitrogen-to-oxygen ratios following the direct method
in the ranges 7.7 $<$ 12+log(O/H) $<$ 8.6 and -1.8 $<$ log(N/O) $<$ -0.8.
They allow us to obtain new empirically calibrated strong-line methods and to evaluate other recipes
based on photoionization models that can be later used for a chemical analysis
of actively star-forming galaxies in very early stages of galaxy evolution.
Our new relations are in agreement with others found for smaller samples of objects at higher redshifts.
When compared with other relations calibrated in the local Universe, they differ when the employed strong-line ratio depends on the hardness of the ionizing radiation, such as O32 or Ne3O2, but they do not when the main dependence is on the ionization parameter, such as S23.
In the case of strong-line ratios depending on [\nii] lines, the derivation of O/H becomes very uncertain due to the
very high N/O values derived in this sample, above all in the low-metallicity regime.
Finally, we adapt the bayesian-like code {\sc HII-Chi-mistry} for the conditions found in this kind of galaxies and we prove that it can be used to derive within errors both O/H and N/O, in  consistency with the direct method .
\end{abstract}




\section{Introduction}

The metal content of the gas in galaxies at different cosmological epochs is one of the
main indicators of their evolution. There are important  scaling relations in galaxies between  metallicity ($Z$) and other
integrated properties  such as stellar mass ($M_*$), known as the mass-metallicity relation (MZR, \citealt{lequeux79, tremonti04}),
or between them in combination with the star formation rate (SFR),
in the known as fundamental metallicity relation (FMR, \citealt{lara-lopez10, mannucci10}).
The evolution of these relations with cosmological time \citep[e.g.][]{lamareille09, cresci10,pm13,zahid13,maier14,kas17}
give important observational constrains to the models of galaxy formation and evolution.

However, one of the main obstacles to study the evolution of these scaling relations is that star-forming galaxies at intermediate and high redshift ($z$) present
very different properties compared with the well-studied
sample of local star-forming regions in disk galaxies. This implies that the methods and the calibrations traditionally used to derive chemical
abundances in  low-$z$ objects must be adapted to the conditions of the high-$z$ samples.

Among the observations pointing out these differences, a direct correlation between $Z$ and the size of galaxies at fixed stellar mass has been found, in the sense
that  smaller radii implies larger $Z$ \citep[e.g.][]{Hoopes07, ellison08,Brisbin12}.
This has implications on the expected metal content of high-$z$ objects as it has been observed that these have in average smaller radii as compared to their low-$z$ analogs for the same stellar mass \citep[e.g.][]{daddi05}. According to simulations, this relation can be pinned down to the relative time when each
galaxy underwent the main episode of gas accretion \citep{jsa18}.

The physical parameters characterizing the nebular emission vary with redshift.
This is the case of the ionization parameter ($U$),
that is known from different local samples of \hii\ regions and galaxies to be lower for higher metal content
\citep[e.g.][]{dopita2000,hcm14}. This dependence can lead
to an enhancement of systematic uncertainties in the derivation of $Z$ and, hence,
to the dispersion in the {\em MZR} found at higher $z$, when the mean $U$ of galaxies is higher \citep{nakajima14,shapley2005, shapley15}.
High-$z$ objects also present higher average electron densities \citep[e.g.][]{brinchmann08, shirazi14,kaasinen17}.
Moreover, the radiation field shows a well defined sequence with redshift in which emission lines in higher-$z$ objects are produced in harder ionizing conditions \citep{kewley13}.

In addition, high-$z$ galaxies can have different chemical evolution histories leading to  abundance ratios
different from  the local sample. This is the case for the nitrogen-to-oxygen ratio (N/O), for  which several authors have found to be higher at a given metallicity  \citep[e.g.][]{masters14,hayashi15,kojima17}, although there is no general consensus on its evolution
 \cite[e.g.][]{shapley15,steidel16,strom18,sanders20}. The difference in behavior could be due to a different star formation efficiency \citep{khochfar11} or the existence of metal-poor gas inflows thought to be responsible for the low O/H and relatively high N/O values observed in Green Pea galaxies \citep{amorin10,amorin12a}
and Lyman-break analogues \citep{loaiza20}. This observed difference can have a non-negligible impact on the determination
of $Z$ using [\nii] emission-lines \citep{pmc09}.

Gas-phase $Z$ in star-forming galaxies are mainly derived by interpreting optical emission-lines from ionized gas, which readily provide the oxygen abundance O/H. Thus, $Z$ and O/H are used interchangeably in literature, as we do here.
 The scale and the uncertainty of the
resulting O/H, depend basically on the detected collisionally excited lines and the method used to
calibrate the relation between the emission-line fluxes and the metal content of the gas.
In the absence of any metal recombination line, usually around 10$^{-4}$ times fainter than \hb\ and hence very difficult to be measured in
high-$z$ galaxies, the most reliable method comes from the determination of the electron temperature and the use of the so-called
direct method \citep[e.g.][]{pm17}. This method depends, however, on the detection of  weak auroral lines, such as for instance
[\oiii] $\lambda$4363 \AA, difficult to detect in weak or metal-rich objects.
In case these lines are too faint, other methods based on more easily observable stronger lines are calibrated to estimate O/H \citep[e.g.][]{pagel79, pmd05, maiolino19}.

In the last years, several attempts have been made to enlarge the sample of high-$z$ galaxies with {\em bona-fide} determination of oxygen abundance using the direct method to establish calibrations consistent with the high-$z$ regime, either using stacking of spectra of local analogs
 to high-$z$ star-forming galaxies \citep[e.g.][]{bian18}, or with the direct acquisition of auroral lines in high-$z$ spectra \citep[e.g.][]{jones15,sanders20}.
Instead, we use in this work a sample of local objects selected from the Sloan Digital Sky Survey (SDSS)
to carry out an extensive study of the chemical properties of
Extreme Emission Line Galaxies (EELG) considered to be analogs to high-$z$ objects.

In the context of providing a calibration for O/H based on strong lines, local EELGs
 present clear advantages with respect to high-$z$  galaxy sets.
The number of available galaxies is much larger, and the signal-to-noise of the individual spectra often allows us to determine, simultaneously,
their O/H and N/O using the direct method. In high-$z$ targets this is possible
only using stacked spectra which may bias the results.
Part of the O/H derivations in our sample of local EELGs are made with the code {\sc Hii-Chi-mistry} (hereafter {\sc HCm}, \citealt{hcm14})
 to complement the empirical calibrations of strong-lines.
This code is based on the use of photoionization models but it results consistent with the direct method and presents two important advantages for the study of
high-$z$ galaxies: i) it allows us to select several different combinations of emission-lines, which is of special relevance for objects observed with the same instrumental setup at different $z$ \citep[e.g.][]{sanders18}, and
ii) it allows us to use [\nii] ] emission lines to determine N/O independently of O/H, reducing the uncertainty owing to the unknown O/H-N/O relation, as may occur in dense environments \citep{edmunds90,kh05}.

The paper is organized as follows: Section~2 presents our sample of selected EELGs including their properties. In Section 3, we describe the derivation of physical properties and chemical abundances following the
direct method in those galaxies with a reliable measurement of [\oiii] $\lambda$4363 \AA.
In Section 4, we provide different empirical calibrations based on strong collisionally excited emission-lines using the abundances derived from
the direct method or from  the direct calibration of the oxygen abundance with the electron temperature. In Section 5, we extend our analysis of the chemical abundances
to the use of the code {\sc HCm}, using appropriate photoionization models to find a solution for O/H, N/O and the
ionization parameter consistent with the direct method. Finally, in Section 6 we summarize our results and present our conclusions.

\section{Description of the EELG sample}

Our EELG sample selection is based on the Automated Spectroscopic K-means-based (ASK) classification
of galaxy spectra (about one-million galaxies with apparent magnitude brighter than 17.8) in the SDSS-DR7 \citep{kmeans}.
The $k-$means clustering algorithm used for the ASK classification is a well-known robust tool, which is
employed in data mining, machine learning and artificial intelligence
\citep[e.g.][]{Everitt1995,bishop2006} and guarantees that similar rest-frame
spectra belong to the same class.
A thorough description and technical details regarding the ASK classification is provided
in \citet{kmeans} and additional properties of the ASK classes inferred from them
can be found in a series of papers \citep{SanchezAlmeida2011,Ascasibar2011,Huertas2011,Aguerri2012,SanchezAlmeida2012}.
Here, we only summarize their main properties in the specific context of our EELG selection.
A more extended discussion on the ASK-based EELG selection will be provided in a forthcoming paper (Amor\'\i n et al, in prep.).

In brief, the ASK classification scheme allows to separate and characterize all the SDSS galaxy
spectra into 28 ASK classes, which are driven only by the shape of the visible spectrum.
As described in \citet{kmeans}, most ($\sim$\,99\%) galaxies in the SDSS-DR7 were classified
into only 17 ASK major classes, with 11 additional minor classes including the remaining $\sim$\,1\%.
All the galaxies in a class have very similar spectra and their average spectrum is the template spectrum of the class.
Both the ASK classification of all galaxies in the SDSS-DR7 and their template spectra are publicly available\footnote{\url{ftp://ask:galaxy@ftp.iac.es/}}\footnote{\url{https://sdc.cab.inta-csic.es/ask/index.jsp}}
through the Spanish Virtual Observatory.

For the present study, we selected galaxies undergoing intense star formation, which are characterised by their high emission line equivalent widths, i.e. EW(\hb) $>$ 30\AA\ and EW([\oiii]) $>$ 100\AA\ \citep[e.g.][]{Amorin2014,amorin2015,Calabro2017}. According to \citet{SanchezAlmeida2012}, these are typically H{\sc ii} galaxies \citep[e.g.][]{Terlevich1991} which belong to a limited number of minor ASK classes, specifically ASK\,15, ASK\,17, ASK\,20, ASK\,21, ASK\,25, ASK\,26 and ASK\,27, and a small number of objects (43) which are outliers of any of the ASK classes defined in
\citet{kmeans}, i.e. they have nearly zero probability to belong to any class.
Their ASK template spectra has two main components (i) a very faint and blue stellar continuum with extremely weak Balmer stellar absorption lines, and (ii) significant nebular continuum and very strong Balmer recombination and collisionally-excited emission lines. The qualitative physical properties of these galaxies based on their ASK template spectra were presented in \citet[][Table~1]{SanchezAlmeida2012}. All of them are metal-poor starbursts dominated by young ($<$\,10\,Myr) stellar populations.

\begin{figure*}
\centering
\includegraphics[width=6cm,clip=]{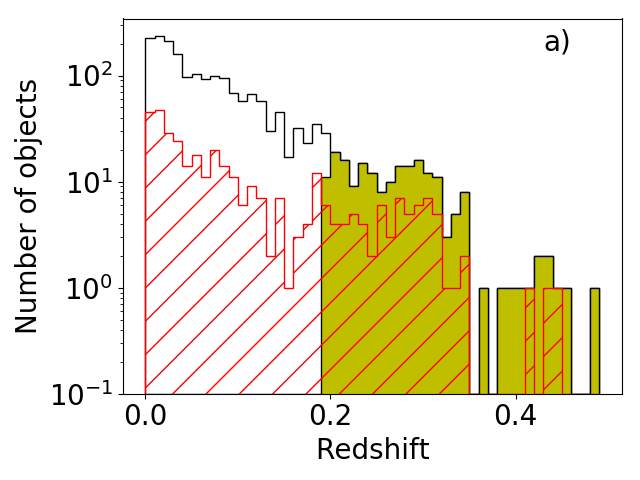}
\includegraphics[width=6cm,clip=]{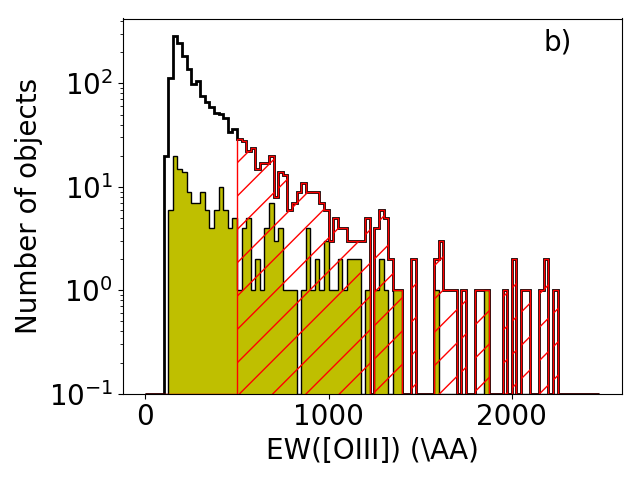}
\includegraphics[width=6cm,clip=]{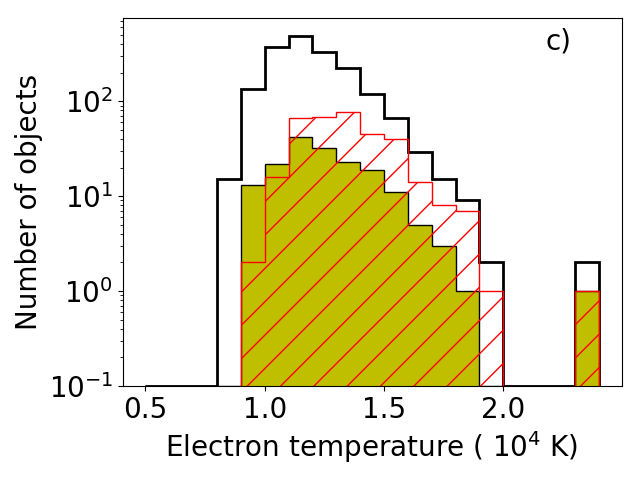}
\includegraphics[width=6cm,clip=]{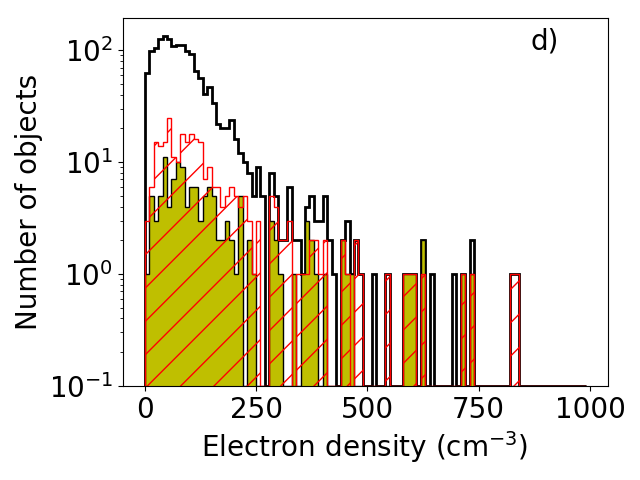}
\includegraphics[width=6cm,clip=]{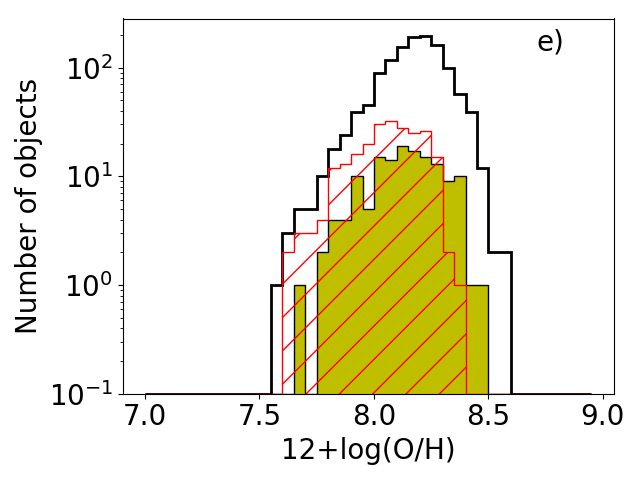}
\includegraphics[width=6cm,clip=]{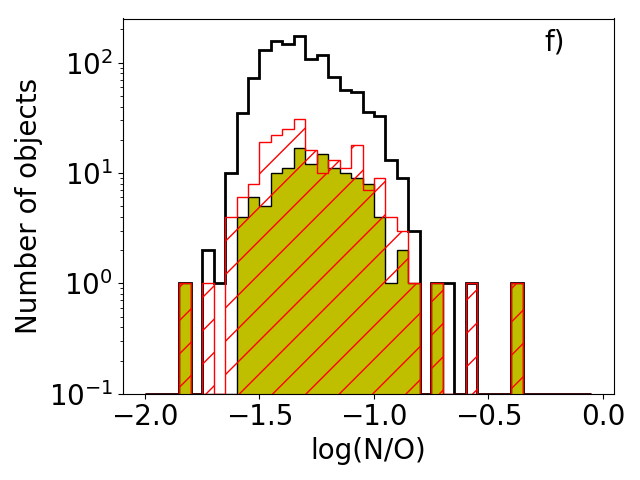}

\caption{Histogram of some of the measured and  derived properties of the selected sample of EELGs, including (a) redshift, (b) equivalent width of [\oiii] 5007 \AA, (c) electron temperature of [\oiii], (d) electron density of [\sii],
(e) total oxygen abundance as derived from the direct method, and (f) nitrogen-to-oxygen ratio, also derived from the direct method.
The yellow histograms show  the objects with a redshift larger than 0.194, i.e., larger
than 90\% of the galaxies in the sample. The red hatched histograms represent objects with EW([\oiii]) $>$ 500 \AA.}

\label{histos}
\end{figure*}

The final sample collects all the galaxies in the ASK clases mentioned above, which all together yield 1969 EELGs with redshifts from $z\sim$\,0 to $z\sim$\,0.49. Nonetheless, the redshift distribution of this sample is not homogeneous and the average $z$ value is 0.08, with a median value of 0.05.
A histogram of this distribution can be seen in upper left panel of Figure \ref{histos}.
The sample includes well-known families of local analogs of high-$z$ galaxies, such as
Green Pea galaxies \citep{cardamone2009,amorin10,amorin12a} and their cohort  \citep{izotov2011}.
After visual inspection, we discarded all duplicated objects within the selected sample.
 The EELG sample includes nearly integrated spectra from extremely compact galaxies such as GPs, and spectra of bright star-forming clumps within larger galaxies.
 For this work, we do not apply any aperture correction on the spectroscopic measurements.

We used the package
{\sc shifu}\footnote{SHerpa IFU line fitting package, (Garc\'ia-Benito in prep.).}
to obtain the flux of the emission
lines from the spectra taken from the Data Release 16 of SDSS \citep{sdss-dr16}. The package contains a suite of routines to easily analyze emission or absorption lines.
The core of the code uses CIAO's Sherpa package
\citep{freeman2001}. Several custom algorithms are implemented in order
to cope with general and ill-defined cases. A first order polynomial was chosen for the continuum, while single gaussians were selected for the lines.

Prior to the emission-line measurement, the spectra
were shifted to the rest frame and corrected for Galactic extinction according to
\cite{Schlegel1998}.
Then, we subtracted the underlying stellar population using the spectral synthesis code {\sc STARLIGHT} \citep{cid2004,cid2005}. {\sc STARLIGHT} fits an observed continuum SED using a nonparametric linear combination of synthesis spectra of different single stellar populations (SSPs), simultaneously solving the ages, metallicities, and the average reddening. In this case, we considered the SSP spectra from \cite{bc2003} based on the {\sc STELIB} library from \cite{LeBorgne2003}, {\sc Padova} 1994 evolutionary tracks, and a \cite{Chabrier2003} initial mass function (IMF) between 0.1 and 100 M$_{\odot}$.
We selected  41 ages spanning from 1 Myr up to 14 Gyr with four metallicities, from $Z$ = 0.0001 up to $Z$ = 0.008.  We used the reddening law from \cite{ccm89} with RV = 3.1.

The emission-line fits
are then performed in the stellar-continuum-subtracted spectra, allowing for the modeling of the continuum to take into account small deviations
in the stellar continuum residuals.
A sigma clipping was independently applied to the residual spectra, and then this was parsed to the composite line plus (residual) continuum model.
Uncertainties in the measured values are evaluated by perturbing the residual spectra according to the error vector 100 times.
Only line fluxes with a signal-to-noise ratio (SNR) larger or equal to 3 were
used for our analysis.
 The continuum was evaluated in the original spectra to determine equivalent widths.
In Figure \ref{histos} (a) we show the resulting histogram of the measured EW for [\oiii] at 5007 \AA. The mean value of the distribution is 354 \AA, with a median value of 249 \AA.
These values are slightly higher when we analyze only the objects at a redshift larger than the 90th percentile of the redshift distribution ($z$ = 0.194) that are, respectively,
of 483 \AA\ and 377 \AA.

In order to carry out our abundance diagnostic analysis, we corrected emission line fluxes from absorption of interstellar dust using the reddening constant
$c(H\beta)$. Its value was obtained from the direct comparison between the observed \ha/\hb\ ratio and the theoretical value for standard physical conditions of the gas ($T_e =$\,10$^4$K
and $n_e=$\,100\,cm$^{-3}$), and assuming the extinction curve of \citet{ccm89}.

We compared the emission-line properties and chemical abundances of our selected EELGs with a local sample of 59\,547 star-forming galaxies, which is used as control sample.
This sample was selected from the parent SDSS-DR7 sample limiting the redshift
range to 0.04\,$<z<$\,0.10 to minimize aperture effects and ensure a S/N of at least 3 in all the relevant strong emission lines used in this work.

\begin{figure}
\centering

\includegraphics[width=8cm,clip=]{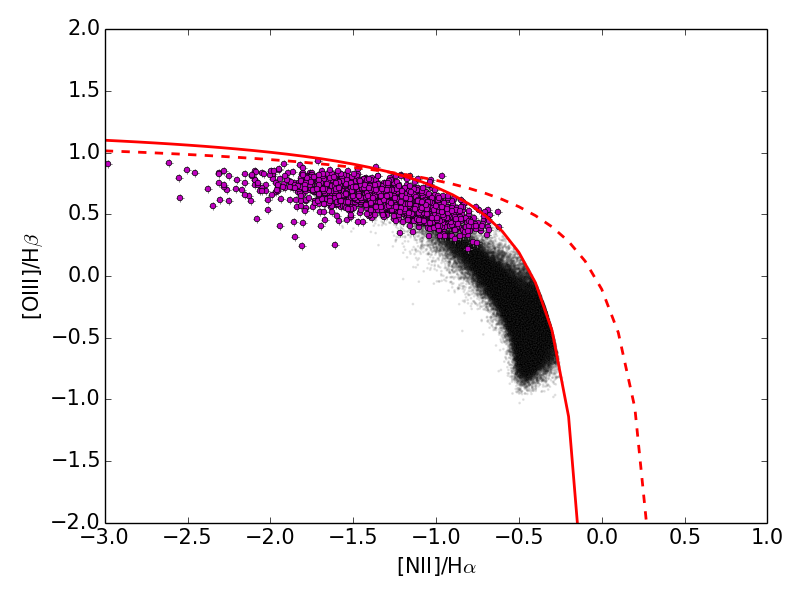}

\caption{Relation between the emission-line ratios [\nii]/\ha\ and [\oiii]/\hb\ for our sample of EELGs and for star-forming
SDSS galaxies. The sample of selected EELGs are symbolized as magenta circles. The
solid line represents the separation curve defined by Kauffmann et al. (2003)  between
star-forming and AGN-dominated galaxies, while the dashed line represents the theoretical separation curve defined by Kewley et al (2001).}

\label{o3-n2}
\end{figure}

In Figure \ref{o3-n2}, we represent the [\nii]/\ha\ vs. [\oiii]/\hb\ diagram,  \citep{bpt},
for both the control sample and the selected EELGs. As it can be seen, both samples  are in regions of this diagram well consistent with emission-lines dominated by stellar photoionization \citep{Kewley2001,Kauffmann2003}.
EELGs present a specific position in this diagram owing to their particular chemical and ionization properties as compared with the rest of star-forming galaxies.
On the one hand, EELGs tend to have larger [\oiii]/H$\beta$ ratios than the
rest (i.e. the mean value of O3 [ = log([\oiii]/\hb)] for EELGs is 0.62
while for the rest of star-forming galaxies is -0.28).
On the other hand, the mean N2 [= log(([\nii]/\ha)] for EELGs (-1.32) is lower
than the control star-forming sample (-0.53).
This difference is significative even if we restrict the star-forming control sample to the
observed [\oiii]/\hb\ range  in EELGs (i.e. $\pm 2\sigma$  the mean value in EELGs).
This makes, as noted in the same figure,
that the EELGs appear close to the separation curve between star-forming
dominated galaxies and AGNs defined by \cite{Kauffmann2003}
and present a very similar position in this diagram to other star-forming galaxies at high-redshift \citep[e.g.][]{bian20}.

\section{Physical properties and chemical abundances from the direct method}

We derived ionic and total chemical abundances of the gas-phase in the selected sample of EELGs using the so-called direct method when the needed emission lines have a S/N larger than 3.
To calculate all the physical properties and chemical abundances described below using these
emission lines,
we employed the analytical expressions listed in \cite{hcm14} and \cite{pm17}, but taking care of using the same
sets of atomic data used in the photoionization models described in the sections below.
This implies using collisional atomic data  from
\cite{Kisielius2009} for O$^+$, and from \cite{Storey2014} for O$^{2+}$.

Briefly, the direct method is based on the determination of the electron temperature
of the ionized gas using the [\oiii] emission-line ratio I(5007)/I(4363).
This ratio could be used for a subsample of 1813 EELGs in our sample, leading to a mean value
of 12\,100 K value, with a standard deviation of 1\,800 K.
Other electron temperatures corresponding to other ions can be measured as well, but the
corresponding auroral line fluxes  are usually weaker than that for of O$^{2+}$.
We checked that none of these auroral lines were neither measurable in the 156 galaxies
for which the [\oiii] auroral line was not measurable.
In those galaxies with just one estimate of electron temperature,  ionic abundances are usually calculated assuming a model-based relation between temperatures.
In this work we use for all the  galaxies with a direct measurement of t([\oiii]),  the model-based relation between t(O$^{2+}$) and t(O$^+$) given by \cite{hcm14} to derive the electron temperatures associated to
the low-excitation regions necessary to calculate their corresponding ionic abundances.

When the temperatures associated to different ionization layers in the gas have been derived, it is possible to
calculate the two main ions abundances (O$^+$ using [\oii] $\lambda$3727 \AA\ and O$^{2+}$ using
[\oiii] $\lambda$5007 \AA). Then the total oxygen abundance is calculated as the addition of
these two abundances relative to H$^+$.
For very high ionization, it is expected that a fraction of oxygen can be ionized to O$^{3+}$, and the use
of an ionization correction factor (ICF) is needed (e.g. P\'erez-Montero et al 2020),
but this is not expected since the high-excitation He{\sc ii} emission line at $\lambda$4686\AA\ is not  observed in most of the objects.
Finally, the nitrogen-to-oxygen ratio is calculated
by means of N$^+$/H$^+$ (via the [\nii] $\lambda$6583 \AA\ line)
and assuming that N$^+$/O$^+$ = N/O.
Both O$^+$ and N$^+$ have non-negligible dependence on electron density,
which was calculated using the values derived from the [\sii] emission-line ratio I(6731)/I(6717). Our sample of EELGs shows gas density values with a mean around 100 cm$^{-3}$
and a standard deviation of 70 cm$^{-3}$, and only in 3 objects we measure electron densities larger than 500 cm$^{-3}$.

The number of objects in our EELGs sample for which it is possible
to derive total oxygen abundances and nitrogen-to-oxygen ratios following the
direct method using the available SDSS spectra are 1268 and 1240 respectively.
These numbers are substantially smaller than the number of galaxies with estimated electron temperature, a drop caused by the fact that the
[\oii] $\lambda$3727 \AA\ emission-line cannot be detected in the most local objects (i.e. $z <$ 0.02) in the observed spectral range of SDSS.
The results that we obtain are in general consistent with those of a family
of metal-poor objects, with
a mean total oxygen abundance from the direct method of 12$+$log(O/H) $=$ 8.17 and
log(N/O)$ = -1.31$, i.e. $\sim$30\% and $\sim$38\% the solar ratios, respectively \citep{asplund09}.
In Figure \ref{histos} we show the distribution of electron temperature as derived from [\oiii], electron density
from [\sii] lines, and chemical abundances from the direct method as discussed above.

To assess the sample homogeneity,  we also show in Figure\ref{histos} the distributions of the same properties for the subsample of EELG with a redshift larger than the 90th percentile of the $z$ distribution in the whole sample (i.e. $z =$ 0.194, with a mean value $z =$ 0.271).
These objects at higher $z$ tend to have slightly higher electron temperatures (12600 K on average) and densities (200 cm$^{-3}$), what translates into slightly lower oxygen abundances (average of 12$+$log(O/H) $=$ 8.13), but a slightly higher  value for log(N/O) ($-1.24$) as compared to the whole sample. These differences can be explained in terms of a larger mean SFR and stellar mass of the objects in the high redshift subsample.
This effect is even more clearly seen in the case of O/H when we restrict our sample to those objects with an EW([\oiii]) of $\lambda$ 5007 \AA\ $>$ 500 \AA, that are also shown in the panels of Figure \ref{histos}.
In this case the mean electron temperature is even larger (13 500 K), the mean
12+log(O/H) snaller (8.05), and N/O has slightly larger mean value ($-1.25$).
The chemical properties of these galaxies  are discussed and compared with other integrated properties, such as M$_*$ or SFR, in Paper II of this series (Amor\'\i n et al, in prep.).

\section{Abundance derivation from empirical calibrations}

\subsection{Relation between O/H and electron temperature}

One of the difficulties to apply the direct method for the
derivation of total chemical abundances is the fact that not all needed lines corresponding
to all ionization stages are available in the optical spectrum.
If the missing ionization stages are not dominant in the total abundance, it is possible to
resort to the use of ICFs based on models
or observations. However, this leads to very high uncertainties when
a large fraction of an element appears in
unobserved stages. This can happen for oxygen
when the [\oii] $\lambda$3727 \AA\ cannot be measured (e.g. when the spectral
coverage does not reach the wavelength region of this line),
which is the case for a subsample of galaxies studied here.

\begin{figure}
\centering

\includegraphics[width=8cm,clip=]{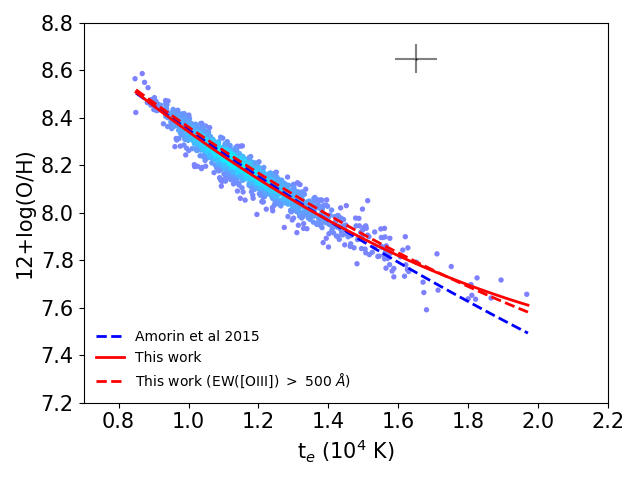}

\caption{Relation between 12+log(O/H) derived from the direct method and the [\oiii] electron temperature for the sample of EELGs where all required lines were available in the SDSS spectra.
The solid red line represents a quadratic polynomial fit to the data, while the dashed blue line represents the linear fit published in
Amor\'\i n et al. (2015).
The red dashed line represents the polynomial fit to the objects with EW([\oiii]) $>$ 500 \AA. The cross in the upper right of the diagram represents the mean errors for both quantities.
}

\label{ohte}
\end{figure}

One alternative can be to measure the weak auroral [\oii] lines at 7319,7330 \AA,
but these do not present an acceptable minimal SNR to be used in these data.
\cite{amorin2015} suggested the use of an empirical relation between
the [\oiii] electron temperature, which only requires the measurement of
[\oiii] $\lambda$
4363 \AA\ and $\lambda$ 4959,5007 \AA, with the total oxygen abundance.
Since this relation was only proposed for local star-forming regions not selected
on the basis of their EW([\oiii]), we study
a new relation fitting the data of our sample of EELGs for which both total oxygen
abundances and [\oiii] electron temperatures were derived.
Taking only those objects with an O/H error lower than 0.2 dex, the
sample consists of 1\,268 galaxies.
The mean [\oiii] electron temperature for this sample is t$_e$([\oiii])$=$11\,800 K
and it approximately ranges from 8\,500 to 19\,700 K.
They are plotted in Fig.~\ref{ohte}. The error-weighted quadratical polynomial

\begin{eqnarray}
\rm
{12+\log(O/H) =} &   {(9.72 \pm 0.04) - (1.70 \pm 0.07) \times t_e +} \nonumber \\
{}  & { + (0.32 \pm 0.03) \times t_e^2}
\end{eqnarray}
with t$_e =$ t$_e$([\oiii]) in units of 10$^4$ K. The standard deviation
of the residuals is 0.04 dex. In Figure \ref{ohte}, this fit
is close to the linear fit derived for a sample of  green-pea galaxies in the Local Universe from \cite{amorin2015}.
A fit considering only objects with EW([\oiii]) $>$ 500 \AA, also shown in the same figure, yields a polynomial very similar to that corresponding to the whole sample.

\subsection{Strong-line methods to derive oxygen abundances}

\begin{figure*}
\centering

\includegraphics[width=8cm,clip=]{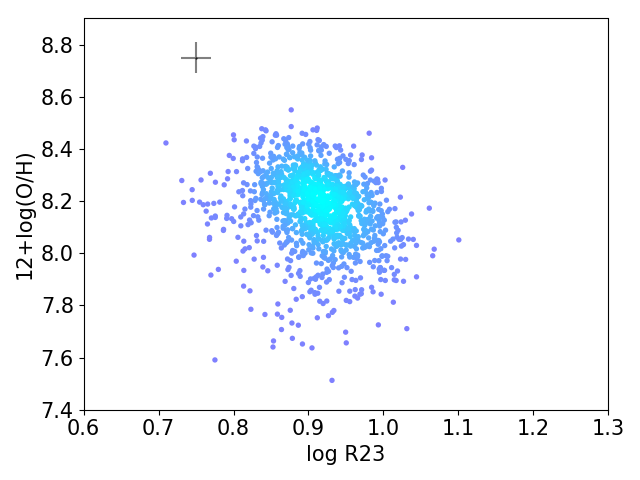}
\includegraphics[width=8cm,clip=]{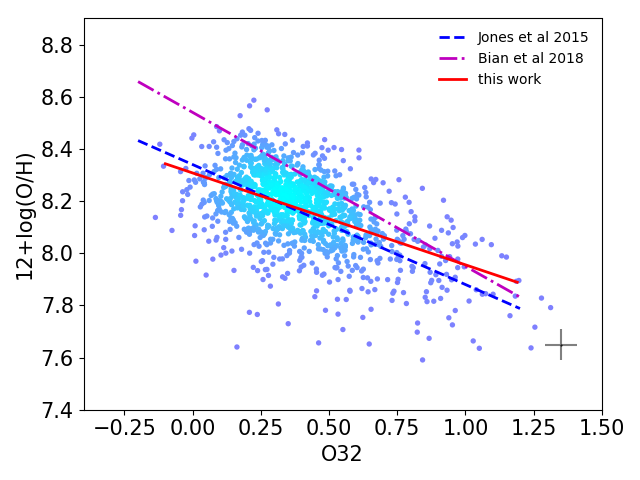}

\caption{Left panel: relation between 12+log(O/H) and log(R23) for those EELGs for which an estimation of O/H based on the electron temperature was possible.
Right panel: Relation with O32 for the same set of EELGs.
The blue dashed line represents the linear fit given by Jones et al. (2015), the
dotted-dashed magenta line represents the fit given by Bian et al (2018), and the
red solid line is a linear fit to the plotted data. In both panels the mean errors are represented with a cross.
}

\label{ohr23}
\end{figure*}

Strong-line methods to derive O/H appear as a common resource when
the direct method or any other prescription based on a previous determination
of the electron temperature cannot be applied. The collisionally excited lines
are usually prominent in the optical spectrum and their relative intensities
depend on the abundances of the ions that produce them.

Among the different strategies that could be followed to calibrate these
strong-line methods, we select only the EELGs  whose chemical abundances were derived following the
direct method, or using a previous estimation of the electron temperature
when the [\oii] line is not measurable (i.e. from eq. 1).
From this subset of EELGs, we only use for the calculation of empirical calibrations those objects that ender O/H with an uncertainty lower than 0.2 dex.

\subsubsection{Parameters based on [OII] and [OIII] emission lines}

One of the most used strong-line parameters to derive O/H is R23, defined
by \cite{pagel79} as,

\begin{equation}
\rm
R23 = \frac{I([OIII] 4959,5007) + I([OII] 3727)}{I(\hb)}
\end{equation}

Among the several limitations of this parameter to provide a precise
determination of the oxygen abundance (see \citealt{pmd05}), it shows a non-negligible dependence on effective temperature, ionization parameter, and it is double-valued with O/H. Indeed R23 increases with increasing O/H at low $Z$ (12+log(O/H $<$ 8.0)
and decreases for increasing O/H at $Z$ (12+log(O/H $>$ 8.4).
In between, in the so-called turn-over region, it is not possible to provide an accurate relation between R23 and O/H.

In Figure~\ref{ohr23}, we show the relation between log(R23) and the
total oxygen abundance for those galaxies in our sample for which an
estimate of the electron temperature could be derived. In this
case all O/H were determined  by means of the direct method as the measurement of the [\oii] emission line at $\lambda$3727 \AA\
was required to compute the behavior of R23. No reliable polynomial
fit can be provided as most of the sample lie
in the turnover region and no clear relation between R23 and O/H exists. Although some galaxies lie either
in the upper or lower branches, these are not enough to provide a calibration.

Other alternative to derive abundances using [\oii] and [\oiii] emission lines is using the ratio between these two lines,

\begin{equation}
\rm
O32 = \log\frac{I([OIII] 4959,5007)}{I([OII] 3727])}
\end{equation}

This ratio is mostly sensitive to ionization parameter and equivalent
effective temperature \citep[e.g.][]{pmd05,dors2003}, but it can also give an estimate
of the total oxygen abundance on the basis of the relation between $Z$ and log U \citep[e.g.][]{dopita2000}.
The right panel of Figure \ref{ohr23} shows the relation between this ratio and 12+log(O/H)
for those local analogs for which it the direct method or Equation~1 could be used
(i.e. 1268 objects).
The correlation is not very marked (i.e. the correlation coefficient, $\rho$ = -0.52) but, as expected, objects with a larger metal content have on average a lower O32 parameter. The red solid line
represents a linear fit to the data that leads to,

\begin{equation}
\rm
12+\log(O/H) = (8.308 \pm 0.007) - (0.352 \pm 0.016) \times O32
\end{equation}
This expression is valid in the range -0.1 $<$ O32 $<$ 1.2,
which corresponds to the range 7.9 $\lesssim$ 12+log(O/H) $\lesssim$ 8.35.
The standard deviation of the residuals in 12+log(O/H) is 0.12 dex.
This fit is similar to the one obtained in \cite{jones15}
(i.e. the data present +0.01 dex mean offset with respect to this calibration)
for a sample of galaxies at intermediate-redshift, showing that our local
sample has on average excitation properties similar to those in other star-forming
galaxies at that redshift range. However, it is sensibly flatter than the linear expression
derived by \cite{bian18} (i.e. the mean offset is -0.14 dex) for a sample of SDSS selected local-analogs calibrated from the
resulting stacked spectra in bins of the N2 parameter.

\subsubsection{Parameters based on [NeIII] emission lines}

\begin{figure*}
\centering

\includegraphics[width=8cm,clip=]{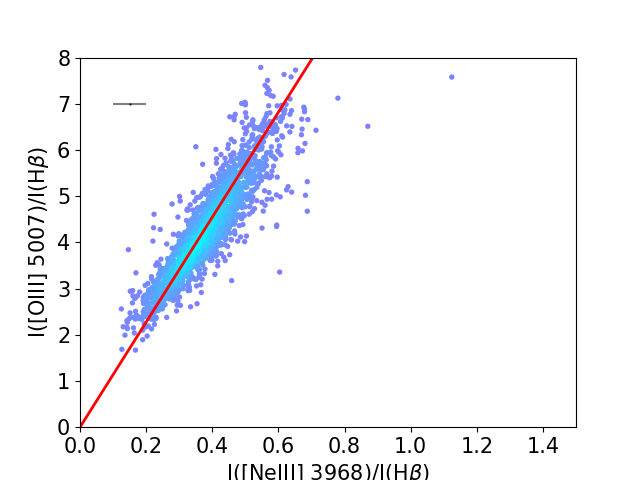}
\includegraphics[width=8cm,clip=]{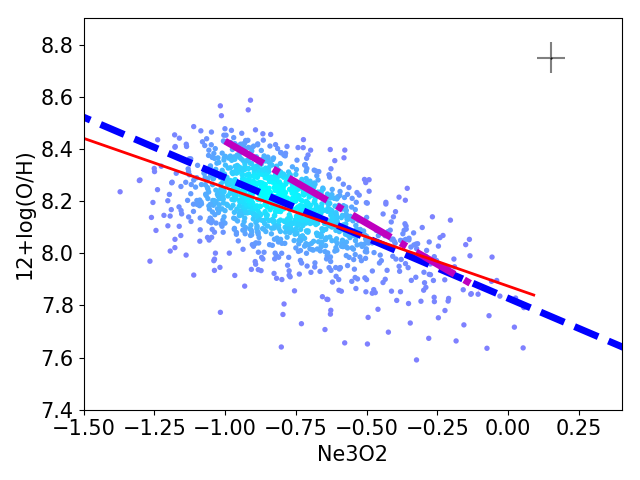}

\caption{Left panel: The relation between the reddening-corrected relative-to-\hb\ intensities of [\oiii] $\lambda$5007 \AA\
and [\neiii] $\lambda$ 3869 \AA\ in our sample of EELGs. Right panel: relation between 12+log(O/H) derived following the direct method and the Ne3O2 parameter. The blue dashed line represents the linear fit given in Jones et al. (2015), the dotted-dashed magenta line represents the fit given by Bian et al (2018), and the
solid red line is a linear fit to the sample analyzed in this work. The crosses in both panels represent the mean errors in abscissae and ordinates.
}

\label{ohne3}
\end{figure*}

In high-$z$ objects, the [\oiii] emission line is often outside the observed spectral range. One alternative to derive O/H
 is using [\neiii]$\lambda$3869 \AA, which is particularly well detected in high excitation objects as those we are studying in the context of this work.

In left panel of Figure \ref{ohne3}, we represent a scatter plot between the
reddening-corrected relative-to-\hb\ intensities of [\oiii] $\lambda$5007 \AA\ and [\neiii]$\lambda$3869 \AA.
As can be seen, these emission lines follow a tight relation (i.e. Spearman rank correlation coefficient $\rho$ = 0.89)
whose slope is $\alpha$ = 11.0 $\pm$ 1.4, that
is slightly lower than the value derived in \cite{pm07} for a sample of local star-forming objects (15.4 $\pm$ 1.5). This tight relation is due to the quite similar excitation structure of O and Ne in the ionized gas and it implies
that all additional dependencies of the [\oiii]/[\oii] emission-line ratio are also present in the
case of [\neiii]/[\oii].

One of these dependencies is the strong relation with ionization parameter that is related with
the metal content of the gas. In this way, we can define the Ne3O2 parameter as

\begin{equation}
\rm
Ne3O2 = \log\frac{I([NeIII] 3869)}{I([OII] 3727])}
\end{equation}

In the right panel of figure \ref{ohne3} we show the relation between this parameter  and12+log(O/H) for those EELGs studied in this work for which this abundance was available through the direct method.
As in the case of O32 the correlation is not very marked ($\rho$ = -0.56), but we can provide a linear fit to this sample that gives,

\begin{equation}
\rm
12+\log(O/H) = (7.823 \pm 0.009) - (0.455 \pm 0.013) \times Ne3O2
\end{equation}
This fit  presents an uncertainty of 0.12 dex, calculated as the standard
deviation of the residuals in the range -1.3 $<$ Ne3O2 $<$ 0.0,
what implies a validity range of 7.8 $\lesssim$ 12+log(O/H) $\lesssim$ 8.4..
As in the case of O32, no significant difference is found with respect to the fit by \cite{jones15} (i.e. the mean
offset with respect to the data is -0.02 dex) for a sample of intermediate-redshift galaxies, although the derived slope
is slightly lower in our case.
Nevertheless, as in the case of O32, a larger difference is found in relation to the linear fit provided by \cite{bian18} (i.e. the mean offset is -0.12 dex) based
on local SDSS analogs, as they found a much pronounced slope for the fit.

\subsubsection{Parameters based on  [NII] emission lines}

\begin{figure*}
\centering

\includegraphics[width=8cm,clip=]{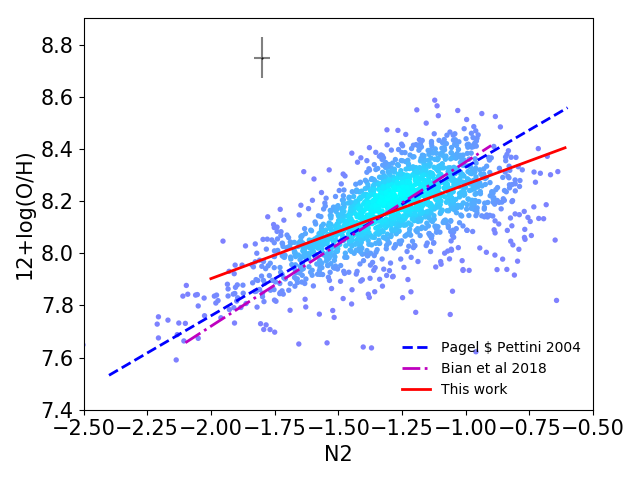}
\includegraphics[width=8cm,clip=]{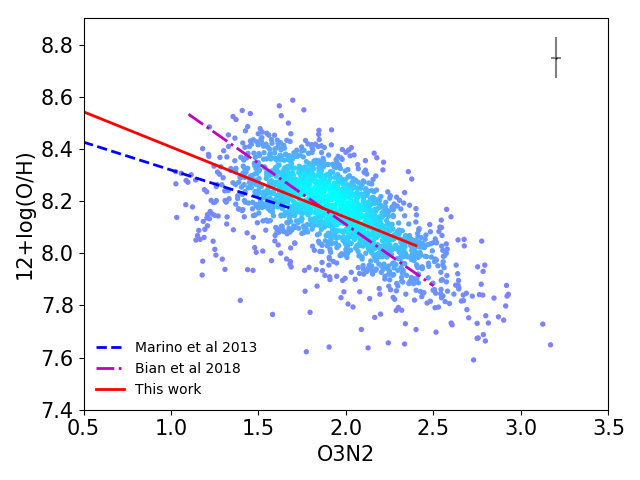}

\caption{Relations between 12+log(O/H) derived from  electron temperature and strong line ratios using [\nii] emission lines fluxes  (N2 in the left panel and O3N2 in the right panel). The solid red line represents linear fits to the plotted data. Note that this fit is only valid for O3N2 $<$ 2.5 in the case of the O3N2 parameter. The dashed blue line represents the linear fit by Pettini \& Pagel (2004) for N2, and the linear calibration
from Marino et al (2013) for O3N2 $<$ 1.8.
The dotted-dashed magenta line represent in both panels the linear fit given by
Bian et al (2018).
In both panels the crosses represent the mean errors.
}

\label{ohn2}
\end{figure*}

Alternatively, other strong-line calibrators based on [\nii] lines have been proposed
to overcome the issue of the double-valued behavior of R23.
The N2 parameter was defined by \cite{storchi} as,

\begin{equation}
{\rm N2} = \log \left( \frac{{\rm I([NII] 6583)}}{{\rm I(H\alpha)}} \right)
\end{equation}
This parameter presents a monotonic relation with O/H and has the advantage
of being easily measurable in the rest-frame red part of the optical spectrum
with very low dependence on reddening or flux calibration.
On the other hand, according to \cite{pmd05} it has a large dispersion due to its
dependence on ionization parameter and N/O abundance ratio.

In  Fig.~\ref{ohn2}, we show the relation between the
N2 parameter and the total oxygen abundance for the 1\,675 objects
of our sample of EELGs for which [\nii] has a SNR $>$ 3  and
there is an estimation of O/H based on the electron
temperature, either from the direct method or from the empirical calibration
with t$_e$ studied above (Eq.~1), and with an uncertainty lower than 0.2 dex.
In this case the correlation is slightly better than for O32 ($\rho$ = 0.66).
However, for metal-poor galaxies (i.e. 12+log(O/H) $<$ 8.0), this correlation worsens. The lack of correlation between N2 and O/H in metal-poor galaxies
has been already reported \cite[e.g.][]{morales14} and it is related to the relatively high N/O ratios found in some of these galaxies. Actually, in our sample log(N/O) does not depend much on metallicity: log(N/O) = -1.34 $\pm$ 0.26  for 12+log(O/H) $<$ 7.9 and  -1.31 $\pm$ 0.16 for 12+log(O/H) $>$ 7.9. A more thorough discussion on this issue will be provided elsewhere (Amor\'\i n et al.,  in preparation).
Thus, we only provide a linear fit in the range  -2.0 $<$ N2 $<$ -0.6
which gives the following expression,

\begin{equation}
\rm
12+\log(O/H) = (8.625 \pm 0.015) + (0.361 \pm 0.012) \times N2
\end{equation}
, which holds for 7.9 $\lesssim$ 12+log(O/H) $\lesssim$ 8.4,
and presents a standard deviation of the residuals of 0.12 dex.
As it can be seen in Figure \ref{ohn2}, the obtained fit gives a less pronounced slope
and a lower intercept
than the linear relation given by \cite{pp04} for a sample of \hii\
regions and \hii\ galaxies in the Local Universe. This implies that the relation for the EELGS
leads to lower values of O/H for the same N2,  in the high-$Z$ regime, even if the mean offset with respect to
the \cite{pp04} calibration is -0.01 dex for this sample.
This  different slope can possibly caused by the different mean ionization parameter
of the two samples. Since log $U$ is much larger in EELGs, low-excitation lines such as [\nii] are fainter for a fixed O/H, so the EELG calibration of this parameter leads to higher values of O/H for the same [\nii] intensities at lower O/H.
This difference is also found in the case of the local SDSS analogs studied in \cite{bian18}, who obtain
a very similar fit to the rest of local \hii\ regions for this parameter and presents a higher slope.

Another combination of collisionally excited strong emission-lines
related to [\nii] $\lambda$6583 \AA\ used to derive O/H is the O3N2 parameter, defined by \cite{alloin} as,

\begin{equation}
\rm
O3N2 = \log \left( \frac{I([OIII] 5007)}{I(\hb)} \cdot \frac{I(\ha)}{I([NII] 6583)} \right)
\end{equation}
This parameter is thought to reduce the dependence on ionization parameter
as it follows the ionization sequence of [\oiii]/\hb\ and [\nii]/\ha\  emission-line ratios
observed in one of the BPT diagrams, as shown in Figure \ref{o3-n2}. The main drawback of using O3N2 is the fact that it does not vary  at low O/H and
according to \cite{pmc09} no clear correlation can be obtained for O3N2 $>$ 2.

The right panel of Fig.~\ref{ohn2} shows the relation between O3N2 and total
oxygen abundance for our sample of EELGs with SNR $>$ 3
in all the required lines and
also show a derivation of the total oxygen abundance based on the electron temperature.
As it can be seen, O/H and O3N2 show a very good correlation in our sample of EELGs, even for large O3N2. Thus, it is possible to provide a linear fit giving (with rho = -0.63),

\begin{equation}
\rm
12+log(O/H) = (8.677 \pm 0.018) - (0.270 \pm 0.009) \times O3N2
\end{equation}
that is obtained in the range 1.0 $<$ O3N2 $<$ 2.7, which corresponds to the range 7.9 $\lesssim$ 12+log(O/H) $\lesssim$ 8.4.
The standard deviation of the residuals is 0.12 dex.
This fit can be compared with the one obtained by \cite{marino13}
for a sample of local \hii\ regions and \hii\ galaxies spanning a  range of O3N2 $<$ 1.8.
In this case, the fit for the EELGS has a slightly larger slope and larger intercept, so this calibration leads to higher values ofO/H for a same value of O3N2 as compared with the calibration for the local star-forming sample (i.e. the mean offset for this calibration is 0.05 dex). In addition, it reaches a larger O3N2 value.
As in the case of N2, this can be interpreted on the basis of a larger average
ionization parameter of the EELGs  sample. Larger values of log U make the O3N2
to be larger for a same O/H, so the calibration leads to larger values of O/H.
Conversely, our fit presents again a smaller slope as compared to that given by \cite{bian18} for their sample of local analogs calibrated using bins of the N2 parameter but, in this case, no
offset is found between the abundances derived using this calibration and the mean value derived for O/H.

\subsubsection{Parameters based on [SII] and [SIII] emission lines}

\begin{figure*}
\centering

\includegraphics[width=8cm,clip=]{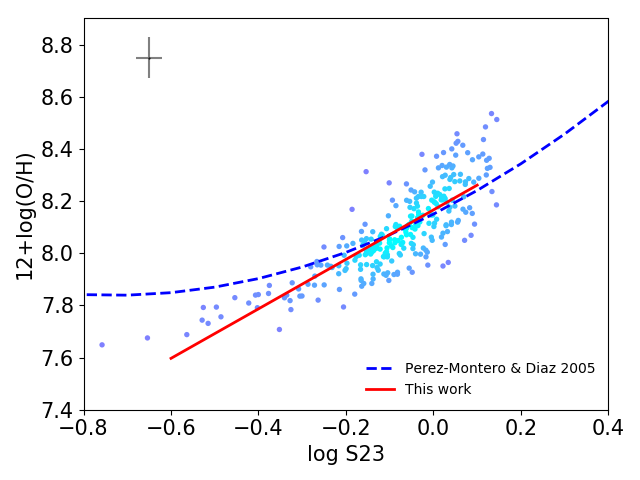}
\includegraphics[width=8cm,clip=]{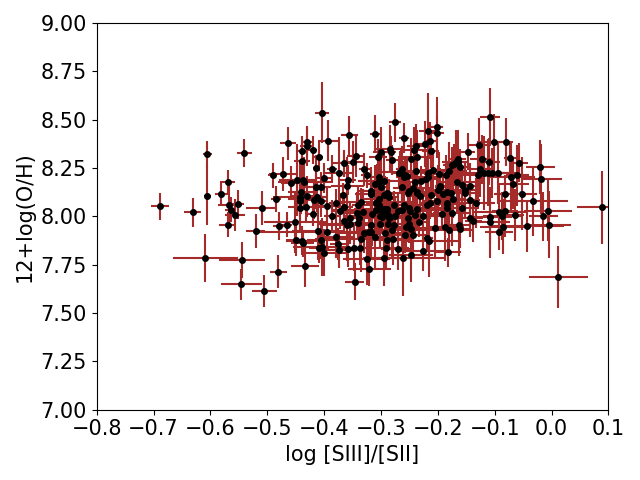}

\caption{Relations between log(S23) (left panel), and log([\siii]/[\sii]) (right panel) with
12+log(O/H) of our EELG sample as derived from a previous estimate of the electron temperature. The solid red line represents a linear fit to the data.
In left panel, the blue dashed line represents the polynomial fit given
by P\'erez-Montero \& D\'\i az (2005).
Crosses represent the mean errors.}

\label{ohs3}
\end{figure*}

\begin{figure}
\centering

\includegraphics[width=8cm,clip=]{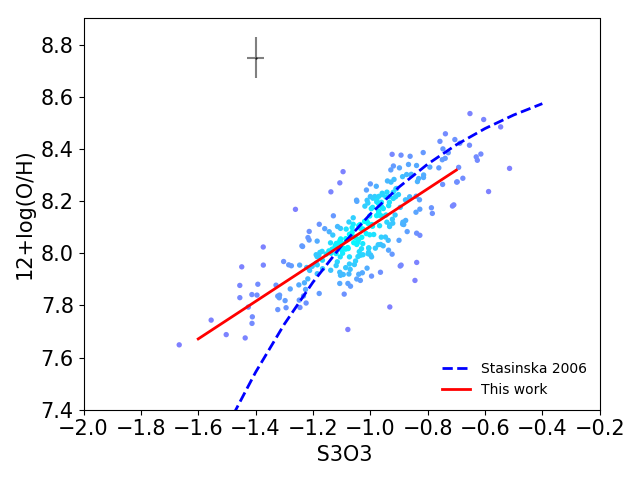}

\caption{Relation between S3O3  with
12+log(O/H) of our EELG sample as derived from a previous estimate of the electron temperature. The solid red line represents a linear fit to the data. The blue dashed line represents the polynomial fit given
by  Stasinska (2006).
The crosses represents the mean errors.}

\label{ohs3o3}
\end{figure}

We also explored strong-line indices sensitive to O/H based on [\sii] and [\siii] lines. The S23 parameter \citep{ve96,dpm2000} is defined as,

\begin{equation}
\rm
S23 = \left( \frac{I([SIII] 9069,9532) + I([SII] 6717,6731)}{I(\hb)} \right).
\end{equation}
This parameter shows a monotonic relation with O/H up to much higher $Z$ than in the case of R23, so it partially fixes the problem of the selection of the appropriate branch in the case of R23. Its dependence on reddening can be overcome using the theoretical relation between \ha\ or \hb\ with the Paschen H{\sc i} recombination lines close to the [\siii] lines. In the case of our EELG sample, the observed SDSS spectra do not cover the redder [\siii] $\lambda$ 9532 \AA\ line, but we can use a theoretical relation with the weaker [\siii] $\lambda$ 9069 \AA\ line (i.e. I([\siii]9532) $=$ 2.44 I([\siii]9069)). On the other hand, even this line is only detected
in the SDSS spectra up to a very low redshift ($z$ $<$ 0.02), and hence  we could only measure it in 349 objects of our sample.
 Moreover, the measurement of the [\siii] $\lambda$ 9069 \AA\ can only be carried out in objects where the [\oii] $\lambda$ 3727 \AA\ lies out of the observed spectral range, so a determination of the total oxygen abundance in these objects is not possible. Alternatively, the empirical calibration
between [\oiii] electron temperature and 12+log(O/H) presented in subsection 4.1 can be used. In left panel of Figure \ref{ohs3}, we show the obtained relation between log S23 and 12+log(O/H) as derived using this method. The resulting correlation is very good ($\rho$ = 0.82), although the O/H range for this subsample  is somehow restricted. A linear fit yields,

\begin{equation}
\rm
12+\log(O/H) = (8.166 \pm 0.07) + (0.949 \pm 0.004) \times log(S23).
\end{equation}
This fit was derived in the range -0.6$<$ log(S23) $<$ 0.2, which corresponds to the range 7.6 $\lesssim$ 12+log(O/H) $\lesssim$ 8.35, and the standard deviation of the
residuals is of 0.09 dex. There are no significant differences with respect to the
quadratical fit by \cite{pmd05} (i.e. the mean offset of the data for this calibration is -0.01 dex) for a sample of local star-forming objects, with the
exception of very low values of O/H.  In our case, a non-linear fit does not
provide a better agreement with the data, as the high-$Z$ range is not properly covered.

As shown in the right panel of Figure \ref{ohs3} and contrary to other high-to-low excitation emission-line
intensity ratios such as O3O2 or Ne3O2,
no correlation between oxygen abundance and the emission line ratio [\siii]/[\sii] is observed ($\rho$ = 0.06).
This different behaviour could be due to the fact that
O32 and Ne3O2 are not directly tracing the excitation of the gas, but the hardness of the ionizing radiation \citep{dors17,pm19a}.
On the contrary, the emission-line ratio [\siii]/[\sii] shows very little dependence on the
equivalent effective temperature of the ionizing source and strongly depends on the ionization parameter \citep[e.g.][]{mathis85,diaz91,morisset16}.
 Thus, the observed relation of O32 or Ne3O2 with O/H could be more  the result of a connection between $Z$ and the hardness of the ionizing source,
 rather than a relation with $U$.

On the other hand, it is possible to derive O/H using the S3O3 parameter, defined as,

\begin{equation}
\rm
S3O3 = \log \left( \frac{I([SIII] 9069,9532)}{I([OIII] 4959,5007)} \right).
\end{equation}
This parameter was proposed as estimator for O/H by \cite{stasinska06}. The relation between  this parameter and
12+log(O/H) is represented in Figure \ref{ohs3o3}.
The correlation in this case is also good ($\rho$ = 0.81) and the linear fit to the data yields the following expression,

\begin{equation}
\rm
12+log(O/H) = (8.822 \pm 0.034) + (0.719 \pm 0.033) \times S3O3.
\end{equation}
The standard deviation of the residuals is only of 0.10 dex in the
range -1.6 $<$ S3O3 $<$ -0.6, which corresponds to 7.6 $\lesssim$ 12+log(O/H) $\lesssim$ 8.4.
There is a good agreement as compared to the polynomial relation proposed by \cite{stasinska06}
(i.e. the mean offset of the data with this calibration is -0.01 dex).

A possible reason for the correlation between this emission-line ratio and O/H is probably the same as that for the cases of O23 or O2Ne3, and contrary to the case of [\siii]/[\sii], mostly due to a correlation with the shape of the spectral energy distribution (SED) than to a dependence on excitation.

\subsection{Strong-line methods to derive nitrogen-to-oxygen abundance ratio}

Several different efforts have been made to also provide strong-line calibrators for the derivation of the
nitrogen-to-oxygen abundance ratio \citep[e.g.][]{thurston96,liang06,pilyugin16},
but little has been explored on calibrations based on  high-redshift objects \citep[e.g.][]{strom18}.
According to \cite{pmc09}, N/O can be derived using
collisional strong emission lines combined to form the N2O2 parameter, defined as,

\begin{equation}
\rm
N2O2 = \log \left( \frac{I([NII] 6583)}{I([OII] 3727)} \right).
\end{equation}

\begin{figure*}
\centering

\includegraphics[width=8cm,clip=]{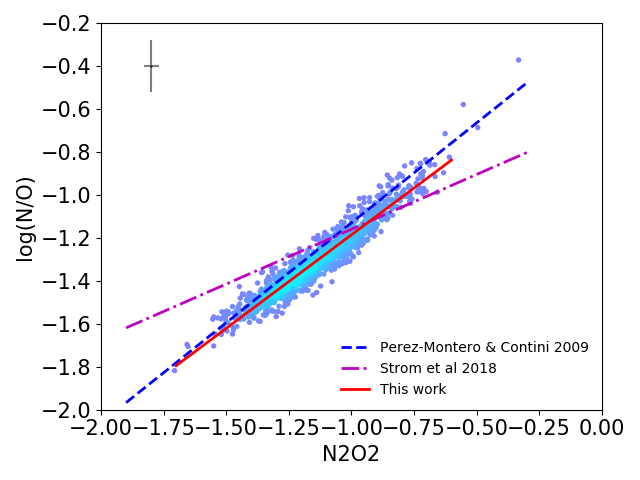}
\includegraphics[width=8cm,clip=]{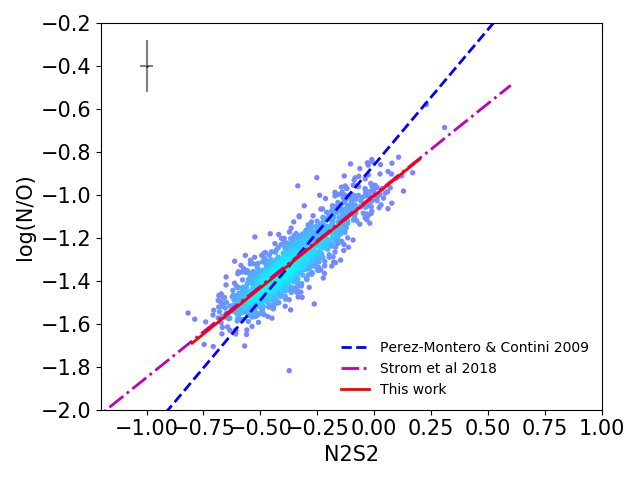}

\caption{Density plots for the relation between log(N/O), as derived following the direct method, and the parameters N2O2 (on the left) and N2S2 (on the right) for the sample of EELGs in our study.
The blue dashed line represent the linear fits obtained in P\'erez-Montero \& Contini (2009) and
the magenta pointed-dashed lines represent the linear fits based on models given by Strom et al (2018).The solid red lines correspond to the linear fits to the sample of EELGs.
Crosses represent the mean errors.
}

\label{non2}
\end{figure*}

This parameter has been also suggested as calibrator for O/H \citep{kewley02}, given its very low dependence on $U$, and
the expected relation between N and O abundances at high-$Z$, where most of the N production has a secondary origin.
In left panel of Fig. \ref{non2} we show the relation between
this emission-line ratio and log(N/O) as derived following the direct
method for those EELGs whose emission lines could be measured
with enough S/R to yield N/O with an uncertainty lower than 0.2 dex.
The observed points present a very tight correlation ($\rho$ = 0.96) and can be fitted by
a linear relation. The fit gives the following expression,

\begin{equation}
\rm
\log(N/O) = (-0.316 \pm 0.009) + (0.870 \pm 0.007) \times N2O2.
\end{equation}

This fit has been calculated in the range -1.7 $<$ N2O2 $<$ -0.5,
which corresponds to -1.8 $\lesssim$ log(N/O) $\lesssim$ -0.75,
and the standard deviation of the residuals is 0.04 dex.
As compared with the linear relation derived by \cite{pmc09}
for a local sample of star-forming regions, the difference is not larger than the
derived errors (i.e. the mean offset of the data to this calibration is -0.03 dex). Since N2O2 does not depend on ionization parameter
the obtained relations based on this parameter do not vary much even
for families of objects whose mean excitation is quite different.
We also compare our calibration with the model-based linear relation with N/O given for this parameter for high-redshift objects by \cite{strom18}, which shows an smaller slope.

The two lines involved in N2O2 have a significant wavelength difference, which makes it more sensitive to observational issues, like calibration errors or the fact that a single observational setup may not yield them simultaneously. To overcome the problem, \cite{pmc09}
also suggests the use of the N2S2 parameter, defined as,

\begin{equation}
\rm
N2S2 = \log \left(   \frac{I([NII] 6583)}{I([SII] 6717,6731)} \right).
\end{equation}

This has been also suggested as calibrator for O/H in high-redshift galaxies \citep{dopita16}, on the basis of a relation between O and N in
the secondary N production regime.
The right panel of Fig. \ref{non2} shows the relation between
this parameter and the nitrogen-to-oxygen ratio as derived following
the direct method
for our sample of EELGs.
The correlation is slightly worse than in the case of N2O2 ($\rho$ = 0.88) and a linear fit to the data gives the following expression,

\begin{equation}
\rm
\log(N/O) = (-1.005 \pm 0.005) + (0.857 \pm 0.013) \times N2S2.
\end{equation}
The fit is valid for the range covered by the data -0.8 $<$ N2S2 $<$ 0.3, which corresponds to the range -1.7 $\lesssim$ log(N/O) $\lesssim$ -0.75.
The standard deviation of the residuals is 0.08 dex.
As in the case of N2O2, there is not large difference compared to the
linear fit derived by \cite{pmc09} for a local sample (there is almost no mean offset), as N2S2 does not present
a strong dependence on ionization parameter.
In addition, differently to the case of N2O2, our linear fit is nearly identical to the model-based one proposed by \cite{strom18} for high-redshift galaxies.

\section{Determination of chemical abundances based on models}

An alternative method to calibrate strong lines to derive chemical abundances is based on the direct comparison with  photoionization models. This approach allows to examine the parameter space in detail avoiding possible observational biases (e.g. the O/H range
or the cut-off in luminosity of the used emission lines).
In this section we explore an adapted version of the code {\sc HCm} \citep{hcm14},
to the conditions of our sample of EELGs and we check the accuracy in the determination of both O/H and N/O as compared to the values derived from the direct method or from the
direct calibration  of the electron temperature. It can be useful for different observed spectral ranges and different emission line sets.

\subsection{Description of the models}

The grid of models used by {\sc HCm} for EELGs was calculated using the photoionization code {\sc Cloudy} v.17.00 \citep{ferland17}.
This code calculates the line fluxes emitted by a gas distribution irradiated by
a central source.
We considered input properties of the models thought to reproduce some of the expected and observed properties
in our sample of EELGs, including synthetic model cluster atmospheres
and gas densities typical in extreme starbursts at low- and intermediate-redshift.
In this case we considered the SEDs from
BPASS v.2.1 \citep{bpass}, assuming an initial mass function
with binaries and a Salpeter slope of $x$ = -1.35 and an upper stellar mass limit of 300 $M_{\odot}$.
The $Z$ of the stars was chosen to match that of the gas as scaled from the total oxygen
abundance, covering a range 12+log(O/H) = [6.9,9.1] in bins of 0.1 dex.
In terms of the solar abundance, the range goes from 1/60 to 2.6 times the solar oxygen abundance in \cite{asplund09}.
A standard dust-to-gas mass ratio was considered in all models, along with a filling factor
of 0.1. Two different electron densities were assumed with values $n_e$ = 100 and 500 particles per cm$^{-3}$,
covering the values found in our sample of EELGs.
The model calculation was stopped when the proportion of electrons in the gas in relation to hydrogen atoms
was lower than 98\%. The resulting geometry of all models was plane-paralell.
To complete the grid, for each O/H  we considered values of log(N/O) in
the range $[-2.0,0.0]$ in bins of 0.125 dex, and values of the ionization parameter
log $U = [-4.0,-1.5]$ in bins of 0.25 dex. This gives 4301 models for each density.

From each model we extracted the emission line fluxes relative to \hb\ relevant for the
calculation of the abundances. These are [\oii] $\lambda$ 3727 \AA,
[\neiii] $\lambda$ 3869 \AA, [\oiii] $\lambda$ 4363, $\lambda$ 5007 \AA, [\nii] $\lambda$ 6584 \AA, and [\sii] $\lambda$ 6717+6731 \AA.

\subsection{Description of the code}

The code {\sc HCm} calculates the total
oxygen abundance, nitrogen-to-oxygen abundance ratio and ionization parameter in  gaseous nebulae ionized by massive stars using optical emission-lines and comparing them using a bayesian-like
method with the predictions from a large grid of photoionization models covering a wide range of input conditions.
The code is described in \cite{hcm14}, and the different later modifications and improvements
appear in version 2 (use of an interpolated grid of models, \citealt{pm16}); version 3 (use of Monte Carlo iterations to
estimate the derived error, \citealt{pm19a}); or version 4 (application for the Narrow Line Region of Seyfert 2 galaxies, \citealt{pm19b}).
Other versions are also devoted to the calculation of the carbon-to-oxygen abundance ratio and the oxygen total abundances using ultraviolet emission lines
(i.e. {\sc HCm-UV}, \citealt{pma17}).
All these versions, along with the one described here for EELGs are
publicly available\footnote{\url{http://www.iaa.csic.es/~epm/HII-CHI-mistry.html}.}.

In brief, the code firstly calculates N/O considering that the emission-line ratios N2O2 and N2S2, described in previous
sections, are almost independent of the ionization parameter. This calculation is performed creating a distribution of $\chi^2$ values with the difference between the
observed emission-line ratios and the prediction of all models in the grid.
The final N/O value is the weighted-mean of this distribution. Its error is
calculated as the quadratical sum of the standard deviation of the same distribution
and the dispersion of a iterative series of Monte Carlo simulations.
For each iteration, the observed emission line fluxes are perturbed using the observed errors.

Once the grid of models is reduced to the calculated N/O value within errors, a new iteration
through the resulting grid constrained in the N/O space is done to calculate O/H and log $U$.
In this case, the code calculates the $\chi^2$ weights using  as observables different
emission-line ratios known to have some dependence on O/H and $U$ (e.g. R23, O32, N2, O3N2). The resulting values and their corresponding errors
are calculated following the same procedure as described above for N/O.

The code allows the use of the default grid of models or of an interpolated grid that enhances
the resolution in the three studied dimensions by a factor of 10. This option minimizes the clustering of the solutions around the values of the grid.
The code also allows to perform calculations attending to different SEDs in the models, including
an instantaneous burst with {\sc POPSTAR} \citep{popstar},
with a Chabrrier IMF and an age of 1 Myr, or a double peak power law for Seyfert 2 galaxies. For this work we use the models described in the above subsection based on
BPASS v. 2.1 \citep{bpass} with binaries, an IMF with upper stellar mass cut-off at 300 M$_{\odot}$ and a slope
of -1.35 at an age of 1 Myr. The  grid with density of 100 cm$^{-3}$ is the one available in the public version the of the code, but
we also comment the results obtained using the higher  density of 500 cm$^{-3}$.

\subsection{Comparison with the direct method}

The chemical abundances for the EELGs obtained with {\sc HCm and the grid described above is compared here with the abundances from the direct method or from electron temperature as parameterized in Eq. (1).}

In Table \ref{disp}, we list the mean errors for O/H and N/O as obtained by the code
 using different combinations of emission lines.
We also give the mean ($\Delta$) and the standard deviation  ($\sigma$) of the residuals as compared with the empirically derived abundances.
These $\Delta$ and $\sigma$ values can be taken as empirical corrections and additional uncertainties, respectively, to the chemical abundances derived by {\sc HCm} when different sets of  emission-lines are used as input. 

\begin{figure*}
\centering

\includegraphics[width=8cm,clip=]{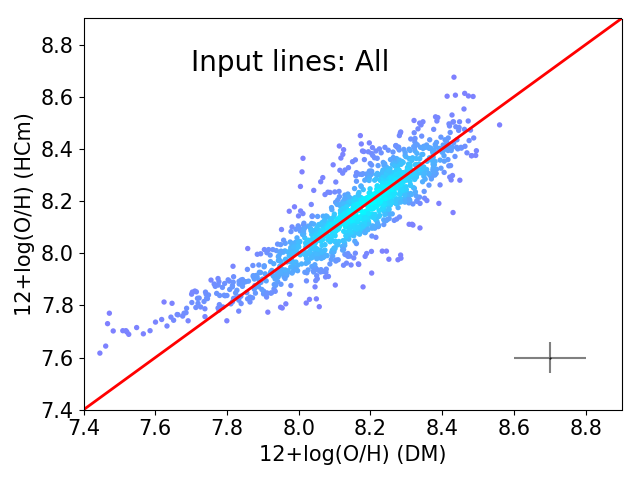}
\includegraphics[width=8cm,clip=]{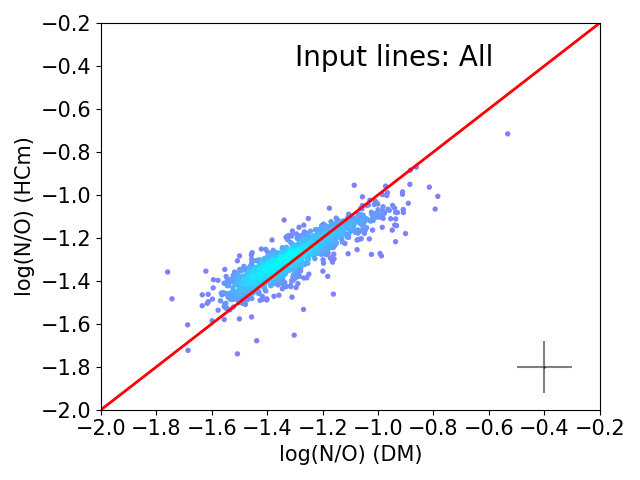}

\caption{Comparison between chemical abundances from the direct method (DM) and
those obtained from {\sc HCm} using all lines for 12$+$log(O/H) (left panel) and
for log(N/O) (right panel). The solid red line represents the one-to-one relation.
Crosses represent the mean errors.}

\label{comp1}
\end{figure*}

Figure \ref{comp1} shows the comparison between the abundances derived following the direct method and
those obtained by our code when all lines are available and the derived abundances have
an uncertainty smaller than 0.2 dex. These criteria include 1269 EELGs.
 The mean offset with respect to the O/H and N/O derived by HCm is in average smaller than the mean errors or the standard deviation of the residuals.
However, for values of 12+log(O/H) $<$ 8.0, the code tends to overestimate the resulting O/H (i.e. 0.1 dex in average). This
offset for this regime was also found in the version of the code for \hii\ regions.
Regarding N/O, for values of log(N/O) lower than -1.6 the code tends to overestimate abundances above 1$\sigma$, but this deviation is reduced when [\sii] is not considered.
We also checked the impact of increasing electron density in the models from 100 to 500 cm$^{-3}$, but we did not found any significant deviation outside the obtained errors for none of the derived quantities.

Contrary to other published abundance determinations based on models  \citep[e.g.][]{kd02,blanc15,asari16}, the agreement between the results obtained from the direct method and from {\sc HCm} is good within the errors.
This difference between the predictions of some models and the abundances derived from the direct method, also known as Abundance Discrepancy Factor (ADF), has been explained invoking
several physical causes, such as shocks \citep[e.g.][]{dopita84,peimbert91}, temperature fluctuations \citep[e.g.][]{peimbert69,garciarojas07}, density or abundance inhomogeneities \citep[e.g.][]{tsamis05}, and $\kappa$ distributions \citep[e.g.][]{dopita2013}.
However,  these physical conditions  should be  introduced
in the models as input conditions to explain the obtained ADF and, if not, models and empirical derivations should
lead to the same results under the same assumptions and atomic data.

The general agreement between the abundances derived from the direct method and from {\sc HCm} also implies that the fraction of unseen oxygen in the optical spectra  (e.g. O$^{3+}$) is negligible in this sample of EELGs.

The comparison between the abundances derived from models with those empirically obtained  can be extended to all the sources with an estimate of electron temperature via the emission-line ratio [\oiii] 5007/4363, even if the [\oii] $\lambda$3727 \AA\
is not detected due to the blue cutoff in the SDSS spectral range.
As it can be seen in Table \ref{disp}, the analysis of the 1714 objects in our sample with an accurate determination of temperature leads to O/H in agreement with the direct method within the mean uncertainty, even when [\oii] $\lambda$ 3727 \AA\ is not taken into account.
Regarding the N/O obtained without [\oii], there is also a general agreement within error, although both the mean uncertainty (0.17 dex) and the mean offset (0.08 dex) are higher than those for the N/O derived using [\oii].

\begin{table*}
\begin{minipage}{180mm}
\begin{center}
\caption{Mean errors, and mean ($\Delta$) and standard deviation ($\sigma$) of the residuals in dex units when comparing 12+log(O/H) and log(N/O) derived from {\sc HCm} and the direct method applied to the studied  EELGs. The rows correspond to the different combinations of extinction-corrected relative-to-\hb\ emission lines used by {\sc HCM}.
The mean errors obtained by {\sc HCm} in each case is also given.
 [\oii] stands for $\lambda$3727 \AA, [\neiii] for $\lambda$3869 \AA, [\oiii]a for $\lambda$4363 \AA, [\oiii]n for $\lambda$5007 \AA,
[\nii] for $\lambda$6583 \AA, and [\sii] for $\lambda$6717+6731 \AA.}

\begin{tabular}{clcccccc}
\hline
\hline
&  Set of input emission lines  &  Mean O/H error   & $\Delta$(O/H) & $\sigma$(O/H) & Mean N/O error  &  $\Delta$(N/O) &  $\sigma$(N/O) \\
\hline
  &  All lines & 0.07  &  +0.00 &  0.08  & 0.13  &   -0.01  &  0.08 \\
&  [\oiii]a, [\oiii]n, [\nii], [\sii] &  0.06  & -0.04  &  0.09 &  0.17  &  +0.08 &   0.08\\
&  [\oii], [\oiii]n, [\nii], [\sii]\footnote{Assuming an empirical relation between O/H and $U$}  & 0.05  &   +0.06 &  0.14 &  0.03  &  +0.06 &   0.11 \\
& [\oii],[\oiii]n, [\nii]$^a$  &  0.05  &  +0.08  &  0.12  &  0.03  &  -0.05  &  0.20  \\
& [\oiii]n, [\nii], [\sii]$^a$  &  0.04  &  +0.06  &  0.15  &  0.01  &  +0.07  &  0.11  \\
& [\nii], [\sii]$^a$  &  0.06  & +0.16   &   0.16     &  0.01  & +0.07 &   0.11  \\
& [\oiii]n, [\nii]$^{a,}$\footnote{Assuming an empirical relation between O/H and N/O}   &  0.03  &  +0.10  &  0.12  &  --  & --  &  --  \\
& [\nii]$^{a,b}$  &  0.03  &  +0.10   &   0.12     &  --  &   -- &   --  \\
& [\oii], [\oiii]n$^{a,b}$  &  0.10   &  +0.09  &  0.14   &  --  &    --    &  --  \\
& [\oii], [\neiii]$^{a,b}$   &  0.05  &   -0.04   &   0.16   &  --  &  --   &  --   \\
\hline
\label{disp}
\end{tabular}
\end{center}
\end{minipage}
\end{table*}

\begin{figure*}
\centering

\includegraphics[width=8cm,clip=]{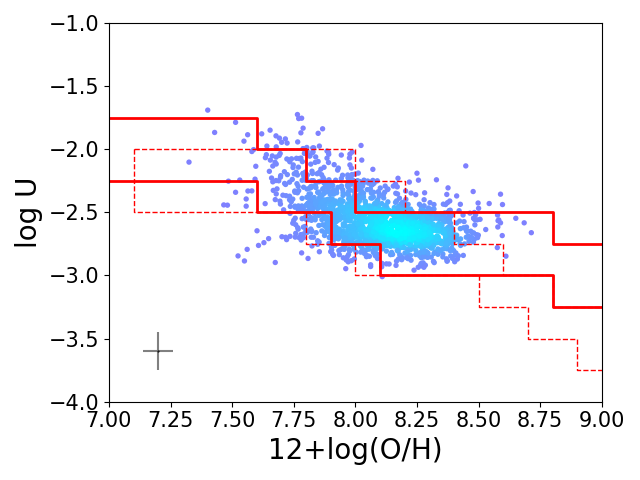}
\includegraphics[width=8cm,clip=]{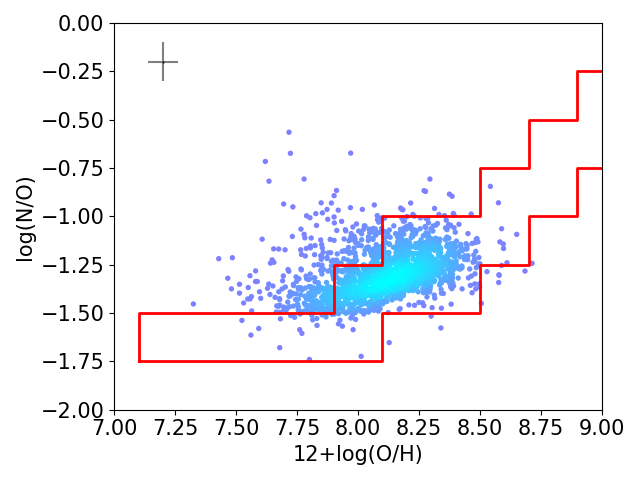}

\caption{Relation between log $U$ (left panel) and log(N/O) (right panel) and 12+log(O/H) calculated using {\sc HCm} for those EELGs having [\oiii] $\lambda$4363 \AA\ emission line. in In both panels, the area encompassed by the solid red lines represent the set of models considered by the code in the absence of [\oiii] 4363 \AA\ or the emision-line ratios N2O2 and N2S2.
The dashed red line in left panel shows the assumption made in the case of the star-forming objects described in P\'erez-Montero (2014). Crosses represent the mean errors.}

\label{OH-U-NO}
\end{figure*}

Since {\sc HCm} gives as result both O/H and $U$, we can explore the
relation between these two properties for the sample of EELGs.
It must be kept in mind that this relation can be also interpreted in terms of an empirical relation between O/H and the hardness of the incident radiation, but our grid of models does not
consider these, as a single SED is assumed for all the models.
The left panel of Figure \ref{OH-U-NO} represents this relation as calculated by {\sc HCm} when the [\oiii] $\lambda$4363 \AA\ emission-line is used.
As in the star-forming regions discussed by \cite{hcm14} there is a trend to find lower excitation in objects with higher O/H, but this is less evident and there is also a mild trend to find metal-rich objects with larger values of $U$.
In fact, $U$ in EELGs tends to be larger than in the rest of star-forming regions.
This can be seen in the same Figure, taking as a reference the grid of models encompassing the local SF regions analyzed by {\sc HCm}. EELGs present log U larger than average, although with significant dispersion.
A new empirical relation between O/H and $U$ can be hence taken as a constraint in the grid of models to derive abundances in absence of the [\oiii] auroral line. In the case of EELGs, this relation adopt values of log $U$ slightly higher than those considered for other star-forming objects at all $Z$ regimes.

\begin{figure*}
\centering

\includegraphics[width=8cm,clip=]{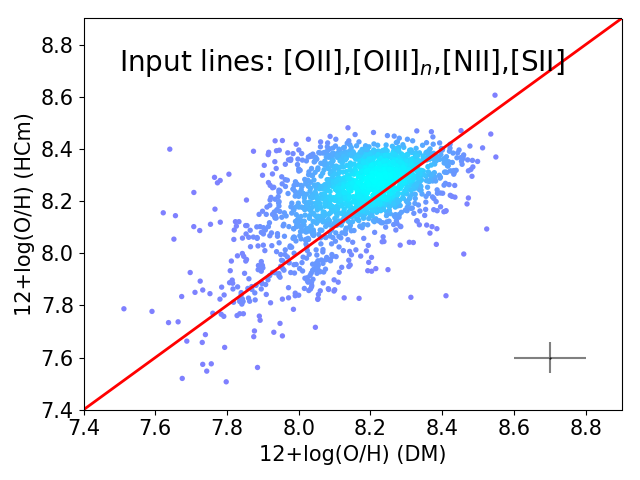}
\includegraphics[width=8cm,clip=]{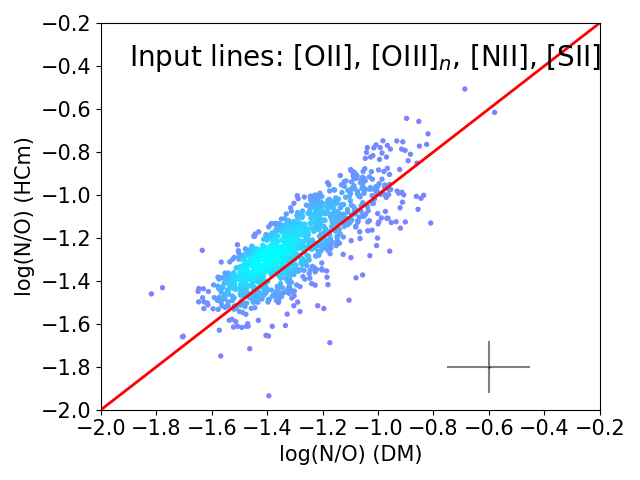}

\caption{Comparison between chemical abundances from the direct method and those obtained from {\sc HCm} using all lines except {\oiii} $\lambda$4363 \AA.
The left and right panels correspond to 12+log(O/H) and log(N/O), respectively.
The solid red line represents the one-to-one relation. Crosses represent the mean errors estimated by {\sc HCm}.}

\label{comp2}
\end{figure*}

Table\ref{disp} also includes the results of the comparison between the abundances derived by the code using a limited set of emission lines in absence of the [\oiii] auroral line and those obtained using the direct method or the empirical relation with the electron temperature.
As shown in Figure \ref{comp2}, the deviations are in general  within the standard deviation of the residuals and, as expected, they increase as the number of used emission lines is smaller. 
Regarding the mean errors derived by the code, the rule is not strict. For example, when all lines are included, the errors are among the largest (Table 1). This can be understood keeping in mind that {\sc HCm} accounts for statistical errors, namely, evaluates how the uncertainties of the fluxes propagate into abundances.
Many sources of systematic error are not included here, and can be significant. Thus, the use of many lines allows bringing out systematic errors bypassed when fewer lines are used and, all in all, may increase {\sc HCm} error when increasing the number of lines.
In any case, as can be seen in Figure \ref{comp2},  {\sc HCm} tends to overestimate the abundances derived from the direct method at low O/H and to underestimate O/H at high $Z$.
The best way to find an accurate determination of the abundance relies always on the detection of the auroral line to derive the electron temperature, however, when this is not possible, {\sc HCm} leads to values consistent at first order with the direct method and allows us to
discriminate, with confidence,  between low- and high-$Z$ objects.

Finally, when no previous determination of N/O can be performed because neither N2O2 nor N2S2 are measured in the observed spectra,
the code has to make a priori assumption about the relation between O/H and N/O in order to use [\nii] lines to derive O/H.
The right panel of Figure~\ref{OH-U-NO} shows the relation adopted in \cite{hcm14} compared to the values obtained in our sample of EELGs.
As it can be seen, many of the sampled EELGs present combinations of N/O and O/H
that lie outside the expected relation, as they show higher N/O values at  low O/H than other low-metallicity samples \citep[e.g.][]{pilyugin12,dopita16,guseva20}.
This plot will be thoroughly discussed in a forthcoming work (Amor\'\i n et al., in preparation), but here it illustrates the risk of assuming  a specified O/H vs N/O  relation for these objects in order to derive
O/H abundances using N emission-lines.

\begin{figure*}
\centering

\includegraphics[width=8cm,clip=]{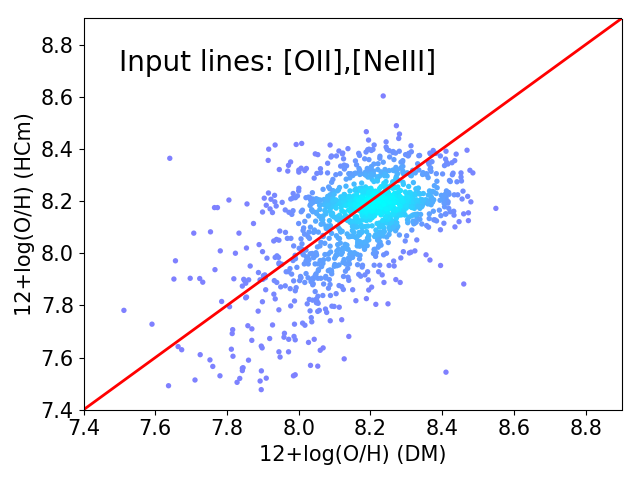}
\includegraphics[width=8cm,clip=]{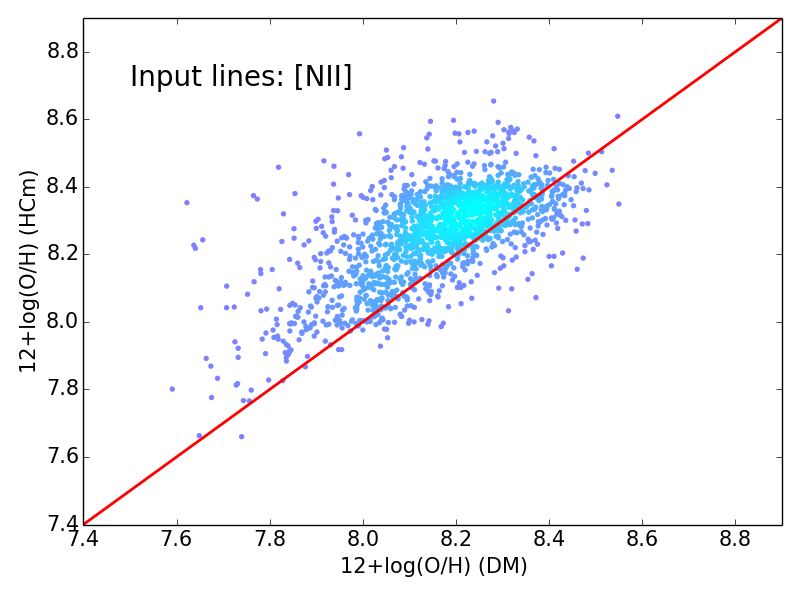}

\caption{Comparison between 12+log(O/H) from the direct method and those obtained from {\sc HCm} using different sets of emission lines: [\oii] and [\neiii], in the left panel, and [\nii] in the right panel, and assuming a relation between N/O and O/O.
The solid red line represents the one-to-one relation. Crosses represent the mean errors estimated by {\sc HCm}.}

\label{comp3}
\end{figure*}

We also show in Table \ref{disp} the values obtained from our analysis when the lines necessary to derive N/O are not available.
As an example, Figure \ref{comp3} compares the O/H derived using {\sc HCm} and the direct method for two combinations of lines where a O/H versus N/O relation has to be assumed. The results based on [\oii] and [\neiii] are slightly better than when only [\nii] is used. In any case, the offset is always at the limit of the standard deviation of the residuals and, as in previous cases, it is sensibly worse in the low-$Z$ regime.

\section{Summary and conclusions}

We presented the analysis of a sample of 1969 EELGs at redshifts $0.00 < z < 0.49$ (mean
$z = 0.08$) selected from the SDSS using the procedure described in \cite{kmeans}.
The sample of EELGs contains compact, high-SFR objects with high \hb\ and [\oiii] equivalent widths. The selected objects  have on average very high values of [\oiii]/\hb\ and low values of [\nii]/\ha, so they are close to the separation curve between star-forming and AGN galaxies in the BPT diagram. The EELG sample gathers analogs of high-$z$ galaxies such as green-pea galaxies \citep{cardamone2009} and their cohort described by \cite{izotov2011}.

Our main goal is to take advantage of the very similar properties of this sample with those of the starbursts at higher-$z$.
We derive their chemical abundances using the direct method and then provide empirical calibrations of the fluxes of their strong emission-lines to be applied to distant objects in the study of the evolution of the main scaling relations involving metallicity such as the MZR or the FMR.

A direct derivation of the total O/H and N/O could be carried out  in around 64\% of the sample, as for $z < 0.02$ the [\oii] 3727 \AA\ line lies outside the observed spectral range in SDSS. The  average O/H turns out to be 0.3 times  the solar value
i.e. the 3$\sigma$ range around the average is 7.7 $<$ 12+log(O/H) $<$ 8.6, while
the average N/O value is 0.36 times the solar value (i.e. covering the range -1.8 $<$ log(N/O) $<$ -0.8). Very small differences are found in the average chemical abundances when we restrict the sample to higher $z$ or EW([\oiii]) values in the sample and these are consistent with an average higher SFR in these subsamples.

For those objects of the studied sample with a good estimation of the electron temperature but for which it is not possible to calculate the total oxygen abundance because [\oii] 3727 \AA\ is not observed, we applied an empirical relation between t([\oiii]) and O/H yielding an uncertainty in 12$+$log(O/H) of only 0.04 dex.

We then explored the empirical calibrations of oxygen abundance for the most commonly used strong-line ratios, in our EELG sample, including O32, Ne3O2, N2, O3N2, S23, and S3O3.
It was not possible to provide calibration based on R23 because our EELGs are in the so-called turnover region where the scatter of R23 with O/H is large.
Regarding O32 or Ne3O2, our linear calibrations present very similar properties to those already provided for intermediate redshift galaxies
\citep[e.g.][]{jones15}.

In the case of strong-line methods based on [\nii] lines, such as N2 or O3N2,
no robust calibration can be given for objects with 12+log(O/H) $<$ 7.8, likely owing to the very high values of N/O
derived in this sample in this regime.
For higher $Z$, our calibrations differ from others based on local samples,  possibly because of differences in the excitation properties.

We also provide empirical calibrations for N/O using the  emission-line ratios N2O2 and N2S2, that lead to good determinations of N/O with uncertainties of less than  0.10 dex.
As neither N2O2 nor N2S2 sensibly depend on $U$, the derived linear calibrations for EELGs are not substantially different from those for other star-forming objects reported by \cite{pmc09}
or the fits based on models for high-redshift galaxies given by \cite{strom18}.

We also investigated the behaviour of other strong-line parameters based on [\siii]
(i.e. S23, S3O3) for a subsample of 335 objects with $z <$ 0.02, and we found that the correlation is much better than that for other parameters.
In any case, no significant difference is found with respect to other calibrations based on other local samples
 \citep[e.g.][]{pmd05,stasinska06}, suggesting that
these calibrations could be based on samples biased towards objects with extreme properties i.e., compatible with those of EELGs.
On the other hand, the lack of correlation between O/H and the emission-line ratio [\siii]/[\sii], which is strongly
dependent on $U$, may indicate that the previously observed relations with other high-to-low excitation emission-line ratios (such as O32 or Ne3O2) results from the dependence of O/H on the hardness of the ionizing radiation rather than on $U$.

Finally, we provide and discuss a version of the code {\sc HII-CHI-mistry} adapted for the studied sample of EELGs using stellar model atmospheres from \cite{bpass}. The agreement between the derived O/H and N/O from the models as compared to those obtained from the direct method are within the errors provided by the code, even in absence of the [\oii] $\lambda$3727 \AA\ line,
although with a slight offset for very low values of $Z$ that must be further investigated.
The $U$ in EELGs derived using {\sc HCm} is slightly larger than the sample of local star-forming
galaxies analyzed in \cite{hcm14}, so the empirical relation between O/H and U used by the code when [\oiii] $\lambda$4363 \AA\ is not detected had to be re-evaluated.
Given the high dispersion shown by this sample in the O/H vs N/O diagram,
a previous determination of N/O to derive oxygen abundances based on [\nii] lines is especially relevant.

The dispersions for different sets of emission-lines are given in Table \ref{disp}
and can be consistently used within errors for different sets of galaxies at different redshift ranges in agreement with the direct method both for O/H and N/O.
A caveat is in order. The error bars provided by {\sc HCm}  are estimative and should be handled with caution since they do not include sources of systematic errors that may be significant.

\section*{Acknowledgements}
We thank the anonymous referee for constructing and helpful comments.
This work has been partly funded by projects Estallidos6 AYA2016-79724-C4 (Spanish Ministerio de Economia y Competitividad), “Estallidos7 PID2019-107408GB-C44
(Spanish Ministerio de Ciencia e Innovacion),
and the Junta de Andaluc\'\i a for grant EXC/2011 FQM-7058.
This work has been also supported by the Spanish Science Ministry "Centro de Excelencia Severo Ochoa Program under grant SEV-2017-0709.
RA acknowledges support from ANID FONDECYT Regular Grant 1202007.
JSA acknowledges support from the Spanish Ministry of Science and
	Innovation, project  PID2019-107408GB-C43 (ESTALLIDOS), and from
	Gobierno de Canarias through EU FEDER funding, project PID2020010050.
RGB acknowledges support from grants PID2019-109067GB-I00 (Spanish Ministerio de Ciencia e Innovaci\'on) and P18-FRJ-2595 (Junta de Andaluc\'\i a).
EPM also acknowledges the assistance from his guide dog Rocko without whose daily help this work would have been much more difficult.
Funding for SDSS, SDSS-II, and SDSS-III has been provided by the Alfred P. Sloan Foundation, the Participating Institutions, the National Science Foundation, and the U.S. Department of Energy Office of Science.
\section*{Data availability}

The data underlying this article are available in the SDSS-DR7 database at
\url{http://classic.sdss.org/dr7/}.
Data products for the EELG sample used in this article will be shared on reasonable
request to the corresponding author.




\bibliographystyle{mnras}
\bibliography{EELG_abs.bib} 








\bsp	
\label{lastpage}
\end{document}